# Inferring energy-composition relationships with Bayesian optimization enhances exploration of inorganic materials

Andrij Vasylenko[1], Benjamin M. Asher[1], Christopher M. Collins[1], Michael W. Gaultois[1], George R. Darling[1], Matthew S. Dyer[1], Matthew J. Rosseinsky[1,*]

[1] Department of Chemistry, University of Liverpool, Crown Street L69 7ZD, UK

*corresponding author

## Abstract

Computational exploration of the compositional spaces of materials can provide guidance for synthetic research and thus accelerate the discovery of novel materials. Most approaches employ high-throughput sampling and focus on reducing the time for energy evaluation for individual compositions, often at the cost of accuracy. Here, we present an alternative approach focusing on effective sampling of the compositional space. The learning algorithm PhaseBO optimizes the stoichiometry of the potential target material while improving the probability of and accelerating its discovery without compromising the accuracy of energy evaluation.

## Main text

A fundamental challenge in materials science is establishing the relationships between the materials' compositions and their synthetic accessibility. For any set of chemical elements (phase field), only a small proportion of viable compositions will afford experimentally accessible phases [1]. There is a global effort to accelerate materials discovery; most approaches focus on reducing the cost of assessment of candidate compositions in high-throughput screening of the phase fields [2–8]. In this Letter, we investigate an alternative approach: can we represent the energy landscape of a phase field as a function of composition (*i.e.*, stoichiometry) and thus accelerate the search for accessible compositions via optimization of such a function?

This hypothesis is motivated by theoretical and experimental observations, in which synthetically accessible phases were realized in the vicinity of computed low-energy compositions with similar stoichiometry [9–14]. Here, we present the learning algorithm PhaseBO that approximates the energy landscape with a simple function and, by selective sampling of the phase field, iteratively improves energy approximation and discovers energy minima.

Computationally, the synthetic accessibility of a composition can be approximated by energy differences with the phases reported in the phase field - the convex hull [9]. There is an increasing number of methods, including density-functional theory (DFT)-based [15–17], interatomic force fields [18–21], and machine learning models [22–26] that aim to improve the accuracy and speed of energy estimation. Most of these approaches use the crystal structure as input; thus, crystal structure prediction (CSP) of a new composition is the most intensive part of its energy evaluation, making exhaustive sampling of a phase field computationally intractable. Composition-based approaches on the other hand offer less reliable energy estimates in comparison to the DFT methods, disfavouring their incorporation into exploratory experimental workflows [27].

In addition to the uncertainties in energy estimation, high-throughput screening methods inherently introduce discretization errors by sampling the phase field. This raises an important question of uncertainty in the approximation of the energy landscape in a compositional space. To address this question, the learning process can incorporate the previous assessments of energy, while taking the uncertainties into account; from the statistical viewpoint, the exploration process should make posterior inference possible, *i.e.*, it should learn to produce the full distribution of possible energies for every point in a continuous compositional space.

In this work, we demonstrate that the energy profile of the compositional space can be approximated as a function of stoichiometry and that this functional dependency can be effectively exploited to accelerate its exploration. Namely, the search for the stable phase – a point on the convex hull – can be approached as an global optimization problem. For this, the energy profile of a compositional space can be approximated with a Gaussian process (GP) and optimized via Bayesian optimization (BO) [28,29]. BO has been proven effective in the exploration of costly-to-evaluate functions and has been increasingly used in materials science to design experiments and optimize sampling [30–46].

Here by employing BO with GP, incorporating previous assessments of energy of compositions into the learning process and taking the uncertainties into account, we enable posterior inference and uncertainty quantification for the whole compositional space, including chemical formulae that are impossible to reliably assess with CSP due to combinatorial complexity. We demonstrate the efficacy of this approach in two examples of previously studied combinatorial spaces Li-Sn-S-Cl [10] and Y-Sr-Ti-O [47], where PhaseBO discovers the experimentally stable and the lowest-energy compositions more consistently and with a higher success probability in comparison to conventional random sampling (RS) and two-step grid search. We also illustrate the capability of the approach to study previously unexplored multi-dimensional compositional spaces, in which high-throughput screening would be extremely costly. In the example of the unexplored Li-Mg-P-Cl-Br phase field, we identify 6 likely candidates for synthetically accessible phases by evaluating only 38 compositions. With demonstrated efficiency in three different solid-state inorganic chemistries, PhaseBO offers routes towards significant acceleration and automation of computationally-driven materials discovery without compromising the accuracy of energy evaluation.

Bayesian optimization represents a class of machine learning methods aiming to find the global optimum in the problem:

$$\min_{x \in \mathbb{R}^d} f(x), \quad (1)$$

where the analytical form of the objective function $f$ in $d$-dimensional space is unknown and its evaluation for any point $x$ is expensive. Here, we represent the search for stable compositions in the materials phase fields as optimization defined in (1). In this formulation, we search for the energy minima above a convex hull, presented as a function of stoichiometry in $d$-element space, in which feasible values of $x$ are stoichiometric coefficients that form a $d$-dimensional simplex $\{x \in \mathbb{R}^d : \sum_i x_i = 1\}$. To find the minima, we employ BO, illustrated in the example quaternary phase field in Figure 1.

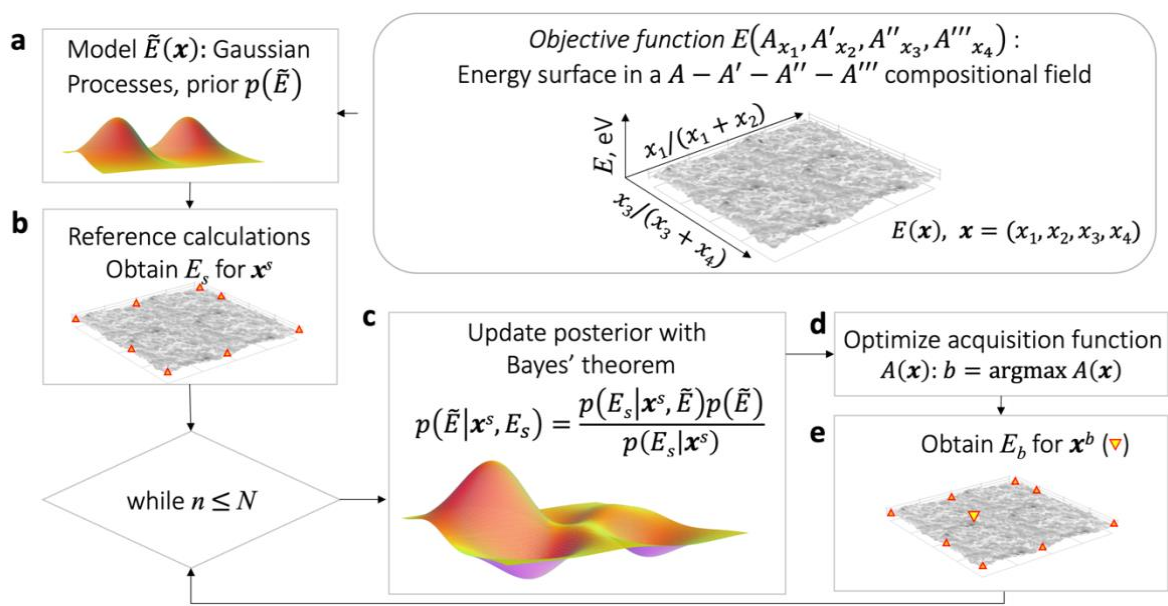

**Figure 1. Schematics of the search for stable compositions – minima of energy $E$ – in a compositional field $A - A' - A'' - A'''$ with Bayesian optimization. a** The true dependence of energy above the

convex hull within the phase field defined by the 4 elements $A, A', A'', A'''$ on compositional content (stoichiometry $x_1, x_2, x_3, x_4$) is the objective function $E(A_{x_1}, A'_{x_2}, A''_{x_3}, A'''_{x_4})$ that we will denote as $E(x), x = (x_1, x_2, x_3, x_4)$, which is unknown and modelled as $\tilde{E}$ with Gaussian processes (GP) with prior probability distribution $p(\tilde{E})$. **b** $E_s = E(x^s)$ energy above the convex hull is calculated for $s$ reference compositions in the phase field (red triangles). **c** The posterior probability distribution $p(\tilde{E} \mid x^s, E_s)$ is updated, its mean represents the model energy function $\tilde{E}$ that best fits the data $\{x^i, E_i\}_{i=1}^{S}$ within this GP. **d** From the posterior, an acquisition function $A(x)$ is built, and its optima suggest compositions $x^b$ for the next evaluation of $E(x^b)$ at stage **e**. Stages **c-e** are repeated while the computational budget allows (number of CSP-studied compositions $n$ is less than a set budget number $N$).

The search for stable compositions and the corresponding minima of energy above the convex hull in the compositional space, *e.g.*, the quaternary $A - A' - A'' - A'''$, starts with the approximation of the unknown objective energy function of stoichiometric coefficients $E(A_{x_1}, A'_{x_2}, A''_{x_3}, A'''_{x_4})$, which we will denote as $E(x)$, $x = (x_1, x_2, x_3, x_4)$ with the GP prior $p(\tilde{E})$ (Figure 1a):

$$p(\tilde{E}) = \mathcal{N}(\mu, k(x, x') + I\sigma_n^2), \quad (2)$$

where the prior is a normal probability distribution of functions $\tilde{E}(x)$, $\mu$ is the mean of the distribution $\mathcal{N}$, $k(x, x')$ is a covariance function, for which we use a Matérn 5/2 kernel [48] (Supplementary Information), $I$ is a unit matrix and the Gaussian noise $\sigma_n^2$ with variance $\sigma_n$ reflects imprecision in energy evaluations.

In BO, information about the objective function is obtained iteratively, following sampling of compositions in the phase field. We start with $s$ reference compositions $x^s$ of a phase field, *e.g.*, in Li-Sn-S-Cl, the references include all reported materials in the phase field (e.g., Li, Sn, S, etc., Supplementary Table 1), whose structures are obtained from databases such as [49], and we evaluate their energy $E_s = E(x^s)$ with DFT (Figure 1 b), that updates the posterior according to Bayes' theorem [28] (Figure 1 c):

$$p(\tilde{E} \mid x^s, E_s) = \frac{p(E_s \mid x^s, \tilde{E}) p(\tilde{E})}{p(E_s \mid x^s)}, \quad (3)$$

where the denominator $p(E_s|\boldsymbol{x}^s)$ is a normalizing constant, optimized with the GP hyperparameters (Supplementary Information).

The posterior mean represents the model energy function $\tilde{E}$ that best fits the data $\mathcal{D}_s = \{\boldsymbol{x}^i, E_i\}_{i=1}^{s}$ in this GP; it can be used to predict energy for unexplored $\boldsymbol{x}$. From the posterior, the acquisition function $A(\boldsymbol{x})$ is constructed (Figure 1 d), which upon optimization, *e.g.*, with gradient-descent, suggests the next composition for evaluation. $A(\boldsymbol{x})$ can be derived in different forms [50] to incorporate a strategy for posterior exploration and exploitation: we employ expected improvement (EI) for sequential and Thompson sampling (TS) [51–53] for batch sampling and compare their performance (Supplementary Information). The posterior update, composition selection and evaluation (Figure 1c – Figure 1e) are repeated until the stopping criteria, such as stable composition discovery or a number of CSP evaluations $N$, defined by a computational budget are satisfied.

To validate this approach, we compare the exploration of a compositional space with conventional methods (random sampling (RS), grid search) and PhaseBO on the examples of two different chemistries, the quaternaries Li-Sn-S-Cl and Y-Sr-Ti-O, which were previously studied with CSP [10,47] (Supplementary Fig.1). Several compositions on the convex hull have been identified in the $Li^+$-$Sn^{4+}$-$S^{2-}$-$Cl^-$ phase field with specified charges, and a new compound ($Li_{3.3}SnS_3Cl_{0.7}$) was discovered experimentally [10] in the vicinity of one of these ($Li_3SnS_3Cl$) ; the lowest-energy composition identified in the $Y^{3+}$-$Sr^{2+}$-$Ti^{4+}$-$O^{2-}$ phase field is 57 meV/atom above the convex hull.

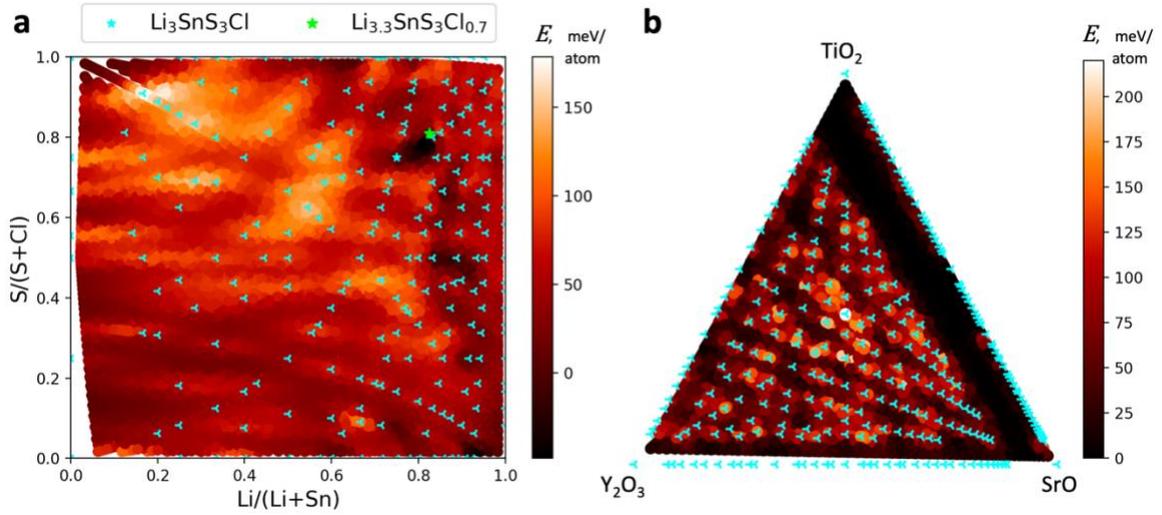

**Figure 2. Exploration of phase fields. a** Li$^+$-Sn$^{4+}$-S$^{2-}$-Cl$^-$, compositions $x^s = \text{Li}_{x_1}\text{Sn}_{x_2}\text{S}_{x_3}\text{Cl}_{x_4}$ presented in 2 dimensions (main text), markers correspond to the total dataset, $s = 195$ compositions with maximum 16 atoms per unit cell, for which energies were calculated with CSP and DFT [10]; the ground state composition Li$_3$SnS$_3$Cl, is marked with a cyan star. The mean posterior calculated with PhaseBO predicts the energy for 49820 unsampled compositions (background colour), including the synthesized Li$_{3.3}$SnS$_3$Cl$_{0.7}$ (lime star). **b** Y$^{3+}$-Sr$^{2+}$-Ti$^{4+}$-O$^{2-}$ is represented in triangular coordinates (main text), markers correspond to the total dataset, $s = 145$ composition with maximum 35 atoms per unit cell, for which energies were calculated with CSP and force field approach in [47]. The mean PhaseBO posterior predicts the energy for 25736 unsampled compositions (background).

In RS, the compositions for evaluation with CSP are selected at random: the sequential choices among accessible compositions are not related. An improved strategy could incorporate results from the previous evaluations, *e.g.,* in a 2-step grid search, where a coarse grid (centroids of clustered compositions) identifies lower energy regions, followed by dense-grid examination of those areas (Supplementary Fig. 2). However, the success of such a strategy depends strongly on the discretization, *i.e.*, on whether a grid dissects a phase field in the vicinity of the low energy compositions. PhaseBO uses previous evaluations to model energy-composition relationships. In Figure 2a, the Li$^+$-Sn$^{4+}$-S$^{2-}$-Cl$^-$ phase field is represented in 2 dimensions: $x' = x_1/(x_1 + x_2)$, $y' = x_3/(x_3 + x_4)$ from the combinations of the stoichiometric coefficients $x_1, x_2, x_3, x_4$ of constituent elements. In Figure 2b, the $Y^{3+}$-$Sr^{2+}$-$Ti^{4+}$-$O^{2-}$ phase field is represented in 2 dimensions: $x' = \frac{1}{2}(x_2 - x_1)$, $y' = \frac{\sqrt{3}}{6}(2x_3 - x_1 - x_2)$, with $x_4$ derived from charge balance. Starting from the reference compositions, for which energies are calculated (Supplementary Table 1-2), the posterior is calculated, and at each iteration, $\text{argmax } A(x)$ suggests further compositions for evaluation. The posterior and uncertainties associated with the GP variance (Supplementary Fig. 3) are updated and the process is repeated until the global minimum is found. The mean posterior can be used to predict the energies for the compositions that cannot or have not yet been assessed with CSP and DFT. In Figure 2, we use the mean posterior to predict the energies for 49820 and 25736 compositions in Li-Sn-S-Cl and Y-Sr-Ti-O, respectively. Importantly, the main features of the energy trends can be reproduced by fitting the energy to only a fraction of all CSP-studied compositions (Supplementary Fig. 4), therefore rapidly highlighting compositional regions of low energy. Thus, the compositional region of the experimentally realized $Li_{3.3}SnS_3Cl_{0.7}$ (lime star in Figure 2) can be discovered, on average, in CSP evaluation for only 40 candidates (Figure 3a, inset).

In Figure 3, we compare the performance of RS, 2-step search and PhaseBO in 300 search experiments. In RS, the probability of randomly discovering a composition depends linearly on the number of accessible compositions, as expected (red dashed line). PhaseBO outperforms conventional approaches in discovering the global minima in Li-Sn-S-Cl and Y-Sr-Ti-O. PhaseBO achieves 100% success of discovery with fewer evaluations, requiring only 50% or less of all compositions.

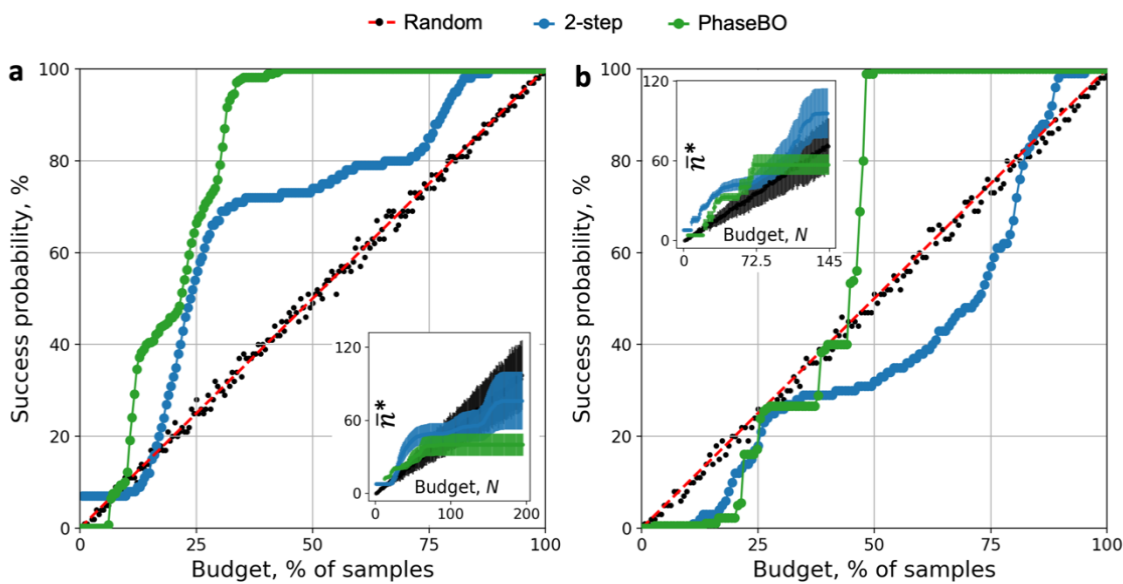

**Figure 3. Performance of conventional and PhaseBO methods for exploration of phase fields.**
**Legend:** Random search (black markers for simulations and dashed line for the trend calculated analytically); 2-step clustering search (blue); PhaseBO (green). **a** Probability of discovering the stable composition ($Li_3SnS_3Cl$) among 195 compositions in Li-Sn-S-Cl is higher with PhaseBO, and reaches 100% after 82 compositions are evaluated (cf 170 for 2-step and 195 randomly). **b** Probability of discovering the lowest-energy composition among 145 compositions in Y-Sr-Ti-O, where PhaseBO reaches 100% after 72 compositions are evaluated (cf 138 for 2-step and 145 randomly). **Insets:** With PhaseBO, the average number of evaluations plateaus at $\bar{n}^* = 40$ compositions for Li-Sn-S-Cl and at 57 compositions for Y-Sr-Ti-O. The shaded areas correspond to standard deviations in 300 experiments.

In the insets of Figure 3, we compare the average number of evaluations, $\bar{n}^*$, required for discovery. PhaseBO requires an average of 40 evaluations to discover the ground state composition in Li-Sn-S-Cl , and on average 57 evaluations in Y-Sr-Ti-O – far fewer evaluations compared to both conventional approaches. Performance of PhaseBO with different acquisition functions, EI presented in Figure3 and TS, is also compared in Supplementary Fig.2, where the advantage of EI is likely due to the larger number of iterations (GP regressions, posterior updates) performed sequentially with EI. This suggests an optimal strategy for the simultaneous exploration of multiple chemistries: via parallelisation over compositional spaces and sequential exploration of individual phase fields. For an unexplored individual compositional space, deceleration of PhaseBO with TS in comparison to EI can be offset against acceleration due to parallel CSP runs, as undertaken below.

We illustrate application of PhaseBO to study an uncharted compositional space, on example of the unexplored Li-Mg-P-Cl-Br phase field, chosen based on synthetic accessibility [10,54] and ionic-conductivity classification [55]. This phase field does not contain any reported quinary phases [49], and makes conventional sampling challenging due to the large combinatorial space. Starting from 8 random compositions, we use USPEX [56] and VASP [57] to generate structures and compare their energies to reference compositions reported in the phase field, constructing a 3-dimensional convex hull. Sampling the posterior with TS of 4 compositions at each iteration, PhaseBO finds 6 compositions with energy above the convex hull < 25 meV/atom in 8 iterations (Figure 4). The lowest-energy compositions found are: $LiMg_2PCl_{10}$ (4 meV/atom) and $LiMg_2PCl_9Br$ (6 meV/atom), which share a similar predicted structure (Supplementary Fig. 5). The full list of the discovered compositions, energies and reference materials are listed in Supplementary Table 1.

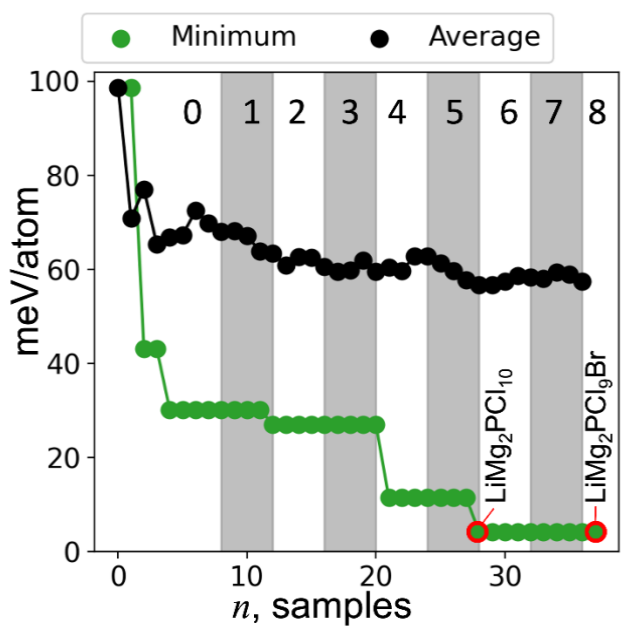

**Figure 4. PhaseBO exploration of the Li-Mg-P-Cl-Br phase field.** PhaseBO discovers LiMg$_2$PCl$_{10}$ (4 meV/atom) and LiMg$_2$PCl$_9$Br (6 meV/atom) – circled red – in 8 iterations, 4 compositions in each – corresponding batch numbers mark grey and white stripes, '0' batch are 8 random compositions. The current minimum and current average energies of sampled compositions (green and black markers, respectively) decrease during PhaseBO exploration.

The sampled energies above the convex hull calculated with CSP (Supplementary Fig. 6) after 8 PhaseBO iterations can suggest promising compositional regions where stable materials may be found.

As demonstrated for 3 CSP methods in 3 different chemistries, PhaseBO coupled with CSP offers a versatile approach to overcome challenges for accurate exploration of compositional spaces at scale, such as intractable computational costs for dense sampling and evaluation of large supercells for modelling disorder. PhaseBO enables rapid identification of low-energy compositional regions, predictions for challenging compositions, and evaluation of uncertainty. It increases the probability of discovery compared to conventional sampling approaches and discovers more potentially attractive candidates more rapidly. It can be applied to other fields of materials discovery, including combinatorial chemistry, where success of the synthesis can be judged by the strength of the diffraction signals. More fundamentally, the successful optimization of energy landscapes suggests a functional relationship between energy and composition, motivating its further examination.

## Acknowledgements

We thank the UK Engineering and Physical Sciences Research Council (EPSRC) for funding through grants number EP/N004884 and EP/V026887. Via our membership of the UK's HEC Materials Chemistry Consortium, which is funded by EPSRC (EP/R029431 and EP/X035859), this work used the UK Materials and Molecular Modelling Hub for computational resources, MMM Hub, which is partially funded by EPSRC (EP/T022213 and EP/W032260)
We thank Dr Andy Zeng for testing the software and for suggestions on the user-friendly features of the PhaseBO.

## Code availability

The software developed for this study is available at

https://www.github.com/lrcfmd/PhaseBO

## Data availability

The data used in this study is available at https://www.github.com/lrcfmd/PhaseBO and available via the University of Liverpool data repository at

https://doi.org/yyy/datacat.liverpool.ac.uk/xxx

## Competing Interests Statement

The authors declare there are no competing interests.

## References

[1] W. Sun, S. T. Dacek, S. P. Ong, G. Hautier, A. Jain, W. D. Richards, A. C. Gamst, K. A. Persson, and G. Ceder, *The Thermodynamic Scale of Inorganic Crystalline Metastability*, Sci. Adv. **2**, e1600225 (2016).

[2] I. Petousis, D. Mrdjenovich, E. Ballouz, M. Liu, D. Winston, W. Chen, T. Graf, T. D. Schladt, K. A. Persson, and F. B. Prinz, *High-Throughput Screening of Inorganic Compounds for the Discovery of Novel Dielectric and Optical Materials*, Sci. Data **4**, 1 (2017).

[3] A. O. Oliynyk, E. Antono, T. D. Sparks, L. Ghadbeigi, M. W. Gaultois, B. Meredig, and A. Mar, *High-Throughput Machine-Learning-Driven Synthesis of Full-Heusler Compounds*, Chem. Mater. **28**, 7324 (2016).

[4] M. de Jong, W. Chen, H. Geerlings, M. Asta, and K. A. Persson, *A Database to Enable Discovery and Design of Piezoelectric Materials*, Sci. Data **2**, 1 (2015).

[5] G. Hautier, A. Miglio, G. Ceder, G.-M. Rignanese, and X. Gonze, *Identification and Design Principles of Low Hole Effective Mass P-Type Transparent Conducting Oxides*, Nat. Commun. **4**, 1 (2013).

[6] D. W. Davies, K. T. Butler, A. J. Jackson, A. Morris, J. M. Frost, J. M. Skelton, and A. Walsh, *Computational Screening of All Stoichiometric Inorganic Materials*, Chem. **1**, 617 (2016).

[7] S. Curtarolo, G. L. W. Hart, M. B. Nardelli, N. Mingo, S. Sanvito, and O. Levy, *The High-Throughput Highway to Computational Materials Design*, Nature Mater. **12**, 3 (2013).

[8] J. E. Saal, S. Kirklin, M. Aykol, B. Meredig, and C. Wolverton, *Materials Design and Discovery with High-Throughput Density Functional Theory: The Open Quantum Materials Database (OQMD)*, JOM **65**, 1501 (2013).

[9] C. Collins, M. S. Dyer, M. J. Pitcher, G. F. S. Whitehead, M. Zanella, P. Mandal, J. B. Claridge, G. R. Darling, and M. J. Rosseinsky, *Accelerated Discovery of Two Crystal Structure Types in a Complex Inorganic Phase Field*, Nature **546**, 7657 (2017).

[10] A. Vasylenko et al., *Element Selection for Crystalline Inorganic Solid Discovery Guided by Unsupervised Machine Learning of Experimentally Explored Chemistry*, Nat. Commun. **12**, 1 (2021).

[11] J. Gamon et al., *Computationally Guided Discovery of the Sulfide $Li_3AlS_3$ in the Li–Al–S Phase Field: Structure and Lithium Conductivity*, Chem. Mater. **31**, 9699 (2019).

[12] E. Shoko, Y. Dang, G. Han, B. B. Duff, M. S. Dyer, L. M. Daniels, R. Chen, F. Blanc, J. B. Claridge, and M. J. Rosseinsky, *Polymorph of LiAlP2O7: Combined Computational, Synthetic, Crystallographic, and Ionic Conductivity Study*, Inorg. Chem. **60**, 14083 (2021).

[13] G. Han et al., *Extended Condensed Ultraphosphate Frameworks with Monovalent Ions Combine Lithium Mobility with High Computed Electrochemical Stability*, J. Am. Chem. Soc. **143**, 18216 (2021).

[14] J. Gamon, M. S. Dyer, B. B. Duff, A. Vasylenko, L. M. Daniels, M. Zanella, M. W. Gaultois, F. Blanc, J. B. Claridge, and M. J. Rosseinsky, *Li4.3AlS3.3Cl0.7: A Sulfide–Chloride Lithium Ion Conductor with Highly Disordered Structure and Increased Conductivity*, Chem. Mater. **33**, 8733 (2021).

[15] A. Wang, R. Kingsbury, M. McDermott, M. Horton, A. Jain, S. P. Ong, S. Dwaraknath, and K. A. Persson, *A Framework for Quantifying Uncertainty in DFT Energy Corrections*, Sci. Rep. **11**, 1 (2021).

[16] E. Sim, S. Song, S. Vuckovic, and K. Burke, *Improving Results by Improving Densities: Density-Corrected Density Functional Theory*, J. Am. Chem. Soc. **144**, 6625 (2022).

[17] V. Blum, R. Gehrke, F. Hanke, P. Havu, V. Havu, X. Ren, K. Reuter, and M. Scheffler, *Ab Initio Molecular Simulations with Numeric Atom-Centered Orbitals*, Comput. Phys. Commun. **180**, 2175 (2009).

[18] Y. Zuo et al., *Performance and Cost Assessment of Machine Learning Interatomic Potentials*, J. Phys. Chem. A **124**, 731 (2020).

[19] C. A. Becker, F. Tavazza, Z. T. Trautt, and R. A. Buarque de Macedo, *Considerations for Choosing and Using Force Fields and Interatomic Potentials in Materials Science and Engineering*, Curr. Opin. Solid State Mater. Sci. **17**, 277 (2013).

[20] J. Behler, *Perspective: Machine Learning Potentials for Atomistic Simulations*, J. Chem. Phys. **145**, 170901 (2016).

[21] S. Batzner, A. Musaelian, L. Sun, M. Geiger, J. P. Mailoa, M. Kornbluth, N. Molinari, T. E. Smidt, and B. Kozinsky, *E(3)-Equivariant Graph Neural Networks for Data-Efficient and Accurate Interatomic Potentials*, Nat. Commun. **13**, 1 (2022).

[22] K. Choudhary and B. DeCost, *Atomistic Line Graph Neural Network for Improved Materials Property Predictions*, NPJ Comput. Mater. **7**, 1 (2021).

[23] C. W. Park and C. Wolverton, *Developing an Improved Crystal Graph Convolutional Neural Network Framework for Accelerated Materials Discovery*, Phys. Rev. Mater. **4**, 063801 (2020).

[24] T. Xie and J. C. Grossman, *Crystal Graph Convolutional Neural Networks for an Accurate and Interpretable Prediction of Material Properties*, Phys. Rev. Lett. **120**, 145301 (2018).

[25] K. T. Schütt, H. E. Sauceda, P.-J. Kindermans, A. Tkatchenko, and K.-R. Müller, *SchNet – A Deep Learning Architecture for Molecules and Materials*, J. Chem. Phys. **148**, 241722 (2018).

[26] C. Chen, W. Ye, Y. Zuo, C. Zheng, and S. P. Ong, *Graph Networks as a Universal Machine Learning Framework for Molecules and Crystals*, Chem. Mater. **31**, 3564 (2019).

[27] C. J. Bartel, A. Trewartha, Q. Wang, A. Dunn, A. Jain, and G. Ceder, *A Critical Examination of Compound Stability Predictions from Machine-Learned Formation Energies*, NPJ Comput. Mater. **6**, 1 (2020).

[28] C. E. Rasmussen, *Gaussian Processes in Machine Learning*, in *Advanced Lectures on Machine Learning: ML Summer Schools 2003, Canberra, Australia, February 2 - 14, 2003, Tübingen, Germany, August 4 - 16, 2003, Revised Lectures*, edited by O. Bousquet, U. von Luxburg, and G. Rätsch (Springer, Berlin, Heidelberg, 2004), pp. 63–71.

[29] B. Shahriari, K. Swersky, Z. Wang, R. P. Adams, and N. de Freitas, *Taking the Human Out of the Loop: A Review of Bayesian Optimization*, Proceedings of the IEEE **104**, 148 (2016).

[30] T. Lookman, P. V. Balachandran, D. Xue, and R. Yuan, *Active Learning in Materials Science with Emphasis on Adaptive Sampling Using Uncertainties for Targeted Design*, NPJ Comput. Mater. **5**, 1 (2019).

[31] A. G. Kusne et al., *On-the-Fly Closed-Loop Materials Discovery via Bayesian Active Learning*, Nat. Commun. **11**, 1 (2020).

[32] A. Solomou, G. Zhao, S. Boluki, J. K. Joy, X. Qian, I. Karaman, R. Arróyave, and D. C. Lagoudas, *Multi-Objective Bayesian Materials Discovery: Application on the Discovery of Precipitation Strengthened NiTi Shape Memory Alloys through Micromechanical Modeling*, Mater. Des. **160**, 810 (2018).

[33] A. Talapatra et al., *Experiment Design Frameworks for Accelerated Discovery of Targeted Materials Across Scales*, Front. Mater. **6**, (2019).

[34] P. I. Frazier and J. Wang, *Bayesian Optimization for Materials Design*, in *Information Science for Materials Discovery and Design*, edited by T. Lookman, F. J. Alexander, and K. Rajan (Springer International Publishing, Cham, 2016), pp. 45–75.

[35] D. Packwood, *Bayesian Optimization for Materials Science*, Vol. 3 (Springer, Singapore, 2017).

[36] T. Ueno, T. D. Rhone, Z. Hou, T. Mizoguchi, and K. Tsuda, *COMBO: An Efficient Bayesian Optimization Library for Materials Science*, Mater. Discov. **4**, 18 (2016).

[37] S. G. Baird, J. R. Hall, and T. D. Sparks, *Compactness Matters: Improving Bayesian Optimization Efficiency of Materials Formulations through Invariant Search Spaces*, Comput. Mater. Sci. **224**, 112134 (2023).

[38] G. Pilania, *Machine Learning in Materials Science: From Explainable Predictions to Autonomous Design*, Comput. Mater. Sci. **193**, 110360 (2021).

[39] F. Häse, L. M. Roch, C. Kreisbeck, and A. Aspuru-Guzik, *Phoenics: A Bayesian Optimizer for Chemistry*, ACS Cent. Sci. **4**, 1134 (2018).

[40] E. O. Pyzer-Knapp, L. Chen, G. M. Day, and A. I. Cooper, *Accelerating Computational Discovery of Porous Solids through Improved Navigation of Energy-Structure-Function Maps*, Sci. Adv. **7**, eabi4763 (2021).

[41] S. Kaappa, C. Larsen, and K. W. Jacobsen, *Atomic Structure Optimization with Machine-Learning Enabled Interpolation between Chemical Elements*, Phys. Rev. Lett. **127**, 166001 (2021).

[42] Q. Liang et al., *Benchmarking the Performance of Bayesian Optimization across Multiple Experimental Materials Science Domains*, NPJ Comput. Mater. **7**, 1 (2021).

[43] A. Seko, A. Togo, H. Hayashi, K. Tsuda, L. Chaput, and I. Tanaka, *Prediction of Low-Thermal-Conductivity Compounds with First-Principles Anharmonic Lattice-Dynamics Calculations and Bayesian Optimization*, Phys. Rev. Lett. **115**, 205901 (2015).

[44] M. Todorović, M. U. Gutmann, J. Corander, and P. Rinke, *Bayesian Inference of Atomistic Structure in Functional Materials*, NPJ Comput. Mater. **5**, 1 (2019).

[45] M. K. Bisbo and B. Hammer, *Efficient Global Structure Optimization with a Machine-Learned Surrogate Model*, Phys. Rev. Lett. **124**, 086102 (2020).

[46] N. Rønne, M.-P. V. Christiansen, A. M. Slavensky, Z. Tang, F. Brix, M. E. Pedersen, M. K. Bisbo, and B. Hammer, *Atomistic Structure Search Using Local Surrogate Model*, J. Chem. Phys. **157**, 174115 (2022).

[47] C. Collins, G. R. Darling, and M. J. Rosseinsky, *The Flexible Unit Structure Engine (FUSE) for Probe Structure-Based Composition Prediction*, Faraday Discuss. **211**, 117 (2018).

[48] D. Cornford, I. T. Nabney, and C. K. I. Williams, *Modelling Frontal Discontinuities in Wind Fields*, J. Nonparametric Stat. **14**, 43 (2002).

[49] D. Zagorac, H. Müller, S. Ruehl, J. Zagorac, and S. Rehme, *Recent Developments in the Inorganic Crystal Structure Database: Theoretical Crystal Structure Data and Related Features*, J. Appl. Cryst. **52**, 5 (2019).

[50] P. I. Frazier, *A Tutorial on Bayesian Optimization*, ArXiv:1807.02811 [Cs, Math, Stat] (2018).

[51] J. M. Hernández-Lobato, J. Requeima, E. O. Pyzer-Knapp, and A. Aspuru-Guzik, *Parallel and Distributed Thompson Sampling for Large-Scale Accelerated Exploration of Chemical Space*, in *Proceedings of the 34th International Conference on Machine Learning* (PMLR, 2017), pp. 1470–1479.

[52] W. R. Thompson, *On the Likelihood That One Unknown Probability Exceeds Another in View of the Evidence of Two Samples*, Biometrika **25**, 285 (1933).

[53] K. Kandasamy, A. Krishnamurthy, J. Schneider, and B. Poczos, *Parallelised Bayesian Optimisation via Thompson Sampling*, in *Proceedings of the Twenty-First International Conference on Artificial Intelligence and Statistics* (PMLR, 2018), pp. 133–142.

[54] A. Vasylenko, D. Antypov, V. Gusev, M. W. Gaultois, M. S. Dyer, and M. J. Rosseinsky, *Element Selection for Functional Materials Discovery by Integrated Machine Learning of Atomic Contributions to Properties*, ArXiv:2202.01051 [Cond-Mat] (2022).

[55] C. J. Hargreaves et al., *A Database of Experimentally Measured Lithium Solid Electrolyte Conductivities Evaluated with Machine Learning*, NPJ Comput. Mater. **9**, 1 (2023).

[56] A. R. Oganov, C. J. Pickard, Q. Zhu, and R. J. Needs, *Structure Prediction Drives Materials Discovery*, Nat. Rev. Mater. **4**, 5 (2019).

[57] G. Kresse and J. Hafner, *Ab Initio Molecular Dynamics for Liquid Metals*, Phys. Rev. B **47**, 558 (1993).